\documentclass[fleqn,usenatbib]{mnras}

\usepackage{newtxtext,newtxmath}
\usepackage[T1]{fontenc}

\DeclareRobustCommand{\VAN}[3]{#2}
\let\VANthebibliography\thebibliography
\def\thebibliography{\DeclareRobustCommand{\VAN}[3]{##3}\VANthebibliography}

\usepackage{graphicx}	% Including figure files
\usepackage{amsmath}	% Advanced maths commands

\usepackage{colortbl}   %Colour
\usepackage{graphicx}   %table size
\usepackage{float}

\newbox\grsign \setbox\grsign=\hbox{$>$} \newdimen\grdimen \grdimen=\ht\grsign
\newbox\simlessbox \newbox\simgreatbox
\setbox\simgreatbox=\hbox{\raise.5ex\hbox{$>$}\llap
     {\lower.5ex\hbox{$\sim$}}}\ht1=\grdimen\dp1=0pt
\setbox\simlessbox=\hbox{\raise.5ex\hbox{$<$}\llap
     {\lower.5ex\hbox{$\sim$}}}\ht2=\grdimen\dp2=0pt
\def\simgreater{\mathrel{\copy\simgreatbox}}

\newbox\simppropto
\setbox\simppropto=\hbox{\raise.5ex\hbox{$\sim$}\llap
     {\lower.5ex\hbox{$\propto$}}}\ht2=\grdimen\dp2=0pt

\title[The origins of N-rich stars in the inner Galaxy]{An enquiry on the origins of N-rich stars in the inner Galaxy based on APOGEE chemical compositions}

\author[S. S. Kisku]{Shobhit Kisku$^{1}$\thanks{E-mail: S.S.Kisku@2015.ljmu.ac.uk},
Ricardo P. Schiavon$^{1}$, Danny Horta$^{1}$, Andrew Mason$^{1}$, 
\newauthor J. Ted Mackereth$^{2,3}$, Sten Hasselquist$^{4,5}$, D. A. Garc\'\i a-Hern\'andez$^{6,7}$, 
\newauthor Dmitry Bizyaev$^{8,9}$, Joel R. Brownstein$^{4}$, Richard R. Lane$^{10}$, Dante Minniti$^{11}$, 
\newauthor Kaike Pan$^{8}$, Alexandre Roman-Lopes$^{12}$
\\
\\
$^{1}$Astrophysics Research Institute, Liverpool John Moores University, 146 Brownlow Hill, Liverpool L3 5RF, UK\\
$^{2}$Canadian Institute for Theoretical Astrophysics, University of Toronto, 60 St. George Street, Toronto, ON M5S 3H8, Canada\\
$^{3}$Dunlap Institute for Astronomy and Astrophysics, University of Toronto, 50 St. George Street, Toronto, ON M5S 3H4, Canada\\ 
$^{4}$Department of Physics and Astronomy, University of Utah, 115 S. 1400 E., Salt Lake City, UT 84112, USA\\
$^{5}$NSF Astronomy and Astrophysics Postdoctoral Fellow\\
$^{6}$Instituto de Astrof\'\i sica de Canarias (IAC), E-38205 La Laguna, Tenerife, Spain\\
$^{7}$Universidad de La Laguna (ULL), Departamento de Astrof\'\i sica, E-38206 La Laguna, Tenerife, Spain\\
$^{8}$Apache Point Observatory and New Mexico State University, P.O. Box 59, Sunspot, NM, 88349-0059, USA\\
$^{9}$Sternberg Astronomical Institute, Moscow State University, Moscow\\
$^{10}$Instituto de Astronom\'\i a y Ciencias Planetarias de Atacama, Universidad de Atacama, Copayapu 485, Copiap\'o, Chile\\
$^{11}$Instituto de Astrof\'\i sica, Pontificia Universidad Cat\'olica de Chile, Av. Vicuna Mackenna 4860, 782-0436 Macul, Santiago, Chile\\
$^{12}$Departamento de F\'\i sica, Facultad de Ciencias, Universidad de La Serena, Cisternas 1200, La Serena, Chile}

\date{Accepted XXX. Received YYY; in original form ZZZ}

\pubyear{2021}

\begin{document}
\label{firstpage}
\pagerange{\pageref{firstpage}--\pageref{lastpage}}
\maketitle

% Abstract of the paper
\begin{abstract}
Recent evidence based on APOGEE data for stars within a few kpc of the Galactic centre suggests that dissolved globular clusters (GCs) contribute significantly to the stellar mass budget of the inner halo.  In this paper we enquire into the origins of tracers of GC dissolution, N-rich stars, that are located in the inner 4 kpc of the Milky Way.  From an analysis of the chemical compositions of these stars we establish that about 30\% of the N-rich stars previously identified in the inner Galaxy may have an accreted origin.   This result is confirmed by an analysis of the kinematic properties of our sample.  The specific frequency of N-rich stars is quite large in the accreted population, exceeding that of its {\it in situ} counterparts by near an order of magnitude, in disagreement with predictions from numerical simulations.  We hope that our numbers provide a useful test to models of GC formation and destruction. 
\end{abstract}

% Select between one and six entries from the list of approved keywords.
% Don't make up new ones.

%%%%%%%%%%%%%%%%%%%%%%%%%%%%%%%%%%%%%%%%%%%%%%%%%%

%%%%%%%%%%%%%%%%% BODY OF PAPER %%%%%%%%%%%%%%%%%%

\section{Introduction}
\label{intro}
One of the main consequences of the current cosmological paradigm, Lambda Cold Dark Matter ($\Lambda$-CDM), is that galaxies grow through the process of hierarchical mass assembly, whereby smaller galaxies are accreted to form larger more massive systems. Such theoretical predictions are in line with the identification of phase-space substructures residing in the Galactic stellar halo, such as Gaia-Enceladus/Sausage (GE/S, \citealp{Belokurov2018, haywood2018, Helmi2018, Mackereth2019}) and Sequoia \citep{Myeong2019}. As well as halo stellar streams (\citealp{Helmi1999, Ibata2016, Belokurov2018}) and ongoing accretion, such as the Sagittarius dwarf spheroidal (Sgr dSph, \citealp{Ibata1994}).
The longer dynamical timescales of less dense regions, such as the outer halo, preserves phase-space information and therefore allows the reconstruction of the integrals of motion (IOM) of these accreted systems.
The situation is not as simple in the inner halo due to the shorter dynamical timescales. Moreover, large extinction towards the inner Galaxy and crowding by more massive metal-rich Galactic components, such as the thick and thin disk, and the bar, make observational access to the inner halo difficult. These difficulties have recently been overcome by the APOGEE survey \citep{APOGEE}, which obtained detailed chemistry based on NIR spectroscopy for over $10^4$ stars in the inner Galaxy, leading up to the discovery of a large population of N-rich stars within a few kpc of the Galactic centre, and the recent identification of Heracles \citep{hortaigs2020}.

In addition to phase-space substructure, stellar streams and ongoing accretion events in the Galactic stellar halo, ancient Globular Clusters (GC) are also thought to contribute relevantly to the total stellar halo mass budget (\citealp{martell2016, Schiavon2017, Koch2019, reina-campos2019, hughes2020, density2020}). Such contribution arises from the dissolution and/or evaporation of GCs, which are disrupted via different processes (\citealp[e.g. tidal shocks, evaporation and disruption by encounters with massive molecular clouds,][]{Gnedin2001, elmegreen2010, kruijssen2011}), so that stars resulting from GC dissolution can be found in the field of the stellar halo.

Detection of the remnants of GC dissolution in the field is made possible by the the presence of stars with chemically peculiar chemical compositions in GCs. These systems have been found to host multiple stellar populations with distinct abundance patterns \citep[for a detailed description, see a review by][]{Bastian-Lardo2018}. Stars that display the same abundances patterns as the field population are dubbed "First Generation" (FG) stars, whereas those that show enhancements in He, N and Na, and show lower O and C are referred to as "Second Generation" (SG) stars. Since abundance patterns of FG stars are indistinguishable from those of field populations, stars with abundance patterns typical of SG population are used as tracers of the contribution of dissolved GCs to the stellar mass budget of the Galaxy.

Field stars that display abundance patterns typical of SG GC stars have been identified in the stellar halo by several groups (\citealp{martellgrebel2010, lind2015, martell2016, Koch2019, tang2019, Tang2020}). Using APOGEE DR12 data, \citet{Schiavon2017} identified a large population of N-rich stars in the inner $\sim$2-3 kpc from the Galactic centre. Based on more recent data releases, these enriched stars have been identified out to large distances up to $\sim$ 15 kpc by \citet{density2020}. 
The large population of N-rich stars identified by \citet{Schiavon2017} is suggested to contribute a minimum of 19-25\% to the stellar mass in the inner $\sim$2 kpc of the halo\footnote{To obtain these numbers, \citet{Schiavon2017} applied the Besançon models \citep{robin2012, robin2014} in order to estimate the contribution of the inner stellar halo to the mass budget of the inner Galaxy.}.
Looking at the halo stars with $|z|>10$ kpc, \citet{martell2016} find the contribution to the stellar mass budget due to GC dissolution to be $\sim2\%$. Such a large spatial variation of the frequency of N-rich stars has been quantified by \citet{density2020}. By taking into account the APOGEE selection effects, they measure a ratio of $\sim17^{+10}_{-7}$\% and $\sim3^{+1}_{-0.8}$\% at R$_{GC}\sim1.5$ kpc and R$_{GC}\sim15$ kpc, respectively.

With the availability of \emph{Gaia's} high-quality parallaxes and the resulting 6D phase-space information, orbital parameters and IOM for Milky Way stars can be estimated. Since these properties are essentially invariant in low density regions of the Milky Way, they can be used to group stars according to orbital properties that are associated to those of the progenitor system.
Recent studies concerning the origins of enriched stars in the halo which show similar abundances to those of SG GCs have investigated the likelihood that these enriched stars originate from GCs (\citealp{Carollo2013, savino&posti2019, Tang2020, hanke2020}).
\citet{savino&posti2019} directly compare the IOM of 57 CN-strong field stars, observed in SEGUE and SEGUE-2 surveys, to those of known Milky Way globular clusters. They find that $\sim$70\% of their sample of field stars have halo-like orbital properties, with only 20 stars having a likely orbital association with an existing globular cluster. They do, however, claim that the orbital properties of halo stars seem to be compatible with the globular cluster escapee scenario.
Similarly, \citet{Tang2020} compare the kinematics of $\sim$100 N-rich stars in LAMOST DR5 to N-normal metal-poor field stars. They conclude that the orbital parameters of N-rich field stars indicate that most of them are inner-halo stars, and that the kinematics of these stars support a possible GC origin.
Note that an alternative way to produce these N-rich stars has been proposed by \citet{Bekki}

In this paper, we aim to constrain the origin of N-rich stars located in the Galactic bulge, on the basis of their chemo-dynamical properties. Identifying a population of accreted and {\it in situ} N-rich stars defined chemically, which are also confirmed by kinematics, we find that the ratio of N-rich to N-normal differ substantially between accreted and {\it in situ} populations.

This paper is organised as follows: In Section \ref{data} we describe the data and the criteria for our sample. The results are presented and discussed in Section \ref{result}, and our conclusions are summarised in Section \ref{sum}.

\section{Data \& Sample}
\label{data}
The results in this paper are based on elemental abundances, radial velocities and stellar parameters from Data Release 16 of the APOGEE-2 survey (\citealp{APOGEE, SDSS-IV, DR16}) and proper motions from \emph{Gaia}-DR2 (\citealp{gaia, gaia2}). We make use of the publicly available code \texttt{galpy}\textit{}\footnote{\url{http://github.com/jobovy/galpy}} (\citealp{Bovy2015, MB2018}) to calculate orbital parameters adopting a \citet{mcmillan2017} potential. We also use distances from \citet{astroNNa} which are generated using the astroNN python package \citep{astroNNb}. The distances are determined using a training set that comprises APOGEE spectra and \emph{Gaia}-DR2 parallax measurements for the purpose of predicting stellar luminosity from spectra. The model is able to simultaneously predict distances and accounts for the parallax offset present in \emph{Gaia}-DR2, producing high precision, accurate distance estimates for APOGEE stars, which match well with external catalogues and standard candles.

\subsection{APOGEE DR16}
\label{apogee}
APOGEE-2, one of the four SDSS-IV (\citealp{SDSS-IV, DR16}) experiments, has obtained near-infrared (NIR), high SNR (S/N > 100 pixel$^{-1}$) and high resolution (R $\sim22,500$) H-band spectra for more than 450,000 Milky Way stars, from which precision radial velocities, stellar parameters, and abundances for up to 26 elements are determined. 
APOGEE-2 uses two twin NIR spectrographs \citep{NIR-spectograph} attached to the 2.5 m telescopes at Apache Point \citep{APO}, and Las Campanas Observatories \citep{lco1973}.
A more in-depth description of the APOGEE survey, target selection, raw data, data reduction and spectral analysis pipelines can be found in \citet{APOGEE}, \citet{Zasowski2017}, \citet{Holtzman2015}, \citet{Jonsson2018}, \citet{Nidever2015}, respectively (see \citet{jonsson2020} for a complete up-to-date description of the latest APOGEE data released in DR16).
The data are first reduced (\citealp{Nidever2015} \& \citealp{jonsson2020}) using the APREAD and APSTAR pipelines, respectively. The data are then fed into the APOGEE Stellar Parameters and Chemical Abundance Pipeline (\citealp[ASPCAP;][]{Garcia-Perez2016, jonsson2020}), which uses libraries of synthetic spectra (\citealp{Zamora2015, Holtzman2018}; and \citealp{jonsson2020}) calculated using customised {\it H}-band line list \citep{shetrone2015}; Smith et al .in prep, from which outputs are analysed, calibrated and tabulated (\citealp{Holtzman2018, jonsson2020}).

\subsection{Sample selection}
\label{sample}
We restrict our sample to stars that have \texttt{ASPCAPFLAG} = 0, 
SNR > 70 and distance uncertainty < 20\% (i.e. \emph{d$_{err}$/\emph{d}} < 0.2). By performing these cuts, we obtain a reduced sample of APOGEE DR16 for which we can obtain reliable chemo-dynamic information.
A further cut of $\log g < 3$ is also made to remove dwarf stars.

In addition, to ensure our sample is free from any stars residing in existing GCs, we remove from our sample any stars belonging to the GC member list from \citet{HortaMNRAS2020}. Furthermore, this paper focuses on stars in the Galactic bulge, so we make a spatial cut and select only stars with Galactocentric distance $R_{GC} <$ 4 kpc. The effective temperature of these stars is further constrained to the range 3250 K $< T_{\rm eff} <$ 4500 K. The lower $T_{\rm eff}$ limit is adopted to avoid very cool stars whose elemental abundances are affected by important systematic effects. The upper limit aims to eliminate from the sample C and N abundances that are uncertain due the weakness of CN and CO lines in spectra of warm stars with relatively low metallicity ([Fe/H] $<-1$).  The bulge selection criteria can be summarised as:
\begin{enumerate}
\item \texttt{ASPCAPFLAG} = 0
\item $R_{GC} <$ 4 kpc
\item \emph{d$_{err}$/\emph{d}} < 0.2
\item 3250 K $< T_{\rm eff} <$ 4500 K 
\item $\log g < 3$
\item SNR > 70
\end{enumerate}

To select our sample of N-rich stars, we follow the sigma clipping methodology implemented in \citet{Schiavon2017}.
By inspecting the bulge stars in the [N/Fe]-[Fe/H] plane, N-rich stars are defined as those deviating by more than $5.5\sigma$ from a 4th order polynomial fit to the data in the bulge sample.
The polynomial is given by: 
\begin{equation}
\begin{split}
{\rm [N/Fe]} \,=\, 0.256 \,+\, 0.239\,{\rm [Fe/H]}\,-\ 0.072\,{\rm [Fe/H]}^2 \\
\,-\ 0.304\,{\rm [Fe/H]}^3\,-\ 0.091\,{\rm [Fe/H]}^4
\end{split}
\label{poly}
\end{equation}
We further restrict these N-rich stars to those with [C/Fe] $<0.15$, in order to limit our sample to stars which present the typical N-C anti-correlation of SG GC stars. Application of these selection criteria leaves us with a sample of 83 N-rich stars within the bulge sample of 14,448 stars.

In this paper we adopt a more stringent threshold of 5.5~$\sigma$ to define N-rich stars than the 4~$\sigma$ threshold adopted by \citet{Schiavon2017}.  In both cases, the threshold decision was informed by the distribution of N-rich stars in abundance planes such as those in Figure~\ref{gcplot}, where N-rich stars display (anti-)correlations between various abundance ratios.  The threshold was chosen so as to clean the N-rich sample from contaminants due to abundance errors and statistical fluctuations.  That philosophy is aimed at prioritising N-rich sample purity over completeness. That our threshold is more stringent than that adopted by \cite{Schiavon2017} reflects the fact that our parent sample is considerably larger, requiring a larger threshold to minimise contamination by outliers due to statistical fluctuations.

We also look at the possible contamination to
our sample of N-rich stars by AGB stars, which can also present an abundance pattern characterised by Nitrogen enrichment and Carbon depletion \citep{Renzini1981, Charbonnel2010, Ventura2013}.
We identified 5 N-rich AGB candidates by their position on the $\log g - T_{\rm eff}$ plane, hand picking those that have low $\log g$, high $T_{\rm eff}$ and relatively high [Fe/H] compared to other stars in their neighbourhood, corresponding to $\sim$6\% of the sample, in agreement with theoretical expectations \citep{girardi2010}. Due to the difficulty of individually selecting AGBs in our large sample of bulge field stars, we decide to keep the N-rich AGBs in our sample for consistency. We note that the results of this paper are largely unaffected by the presence of these N-rich AGBs.

\begin{figure}
	\includegraphics[width=\columnwidth]{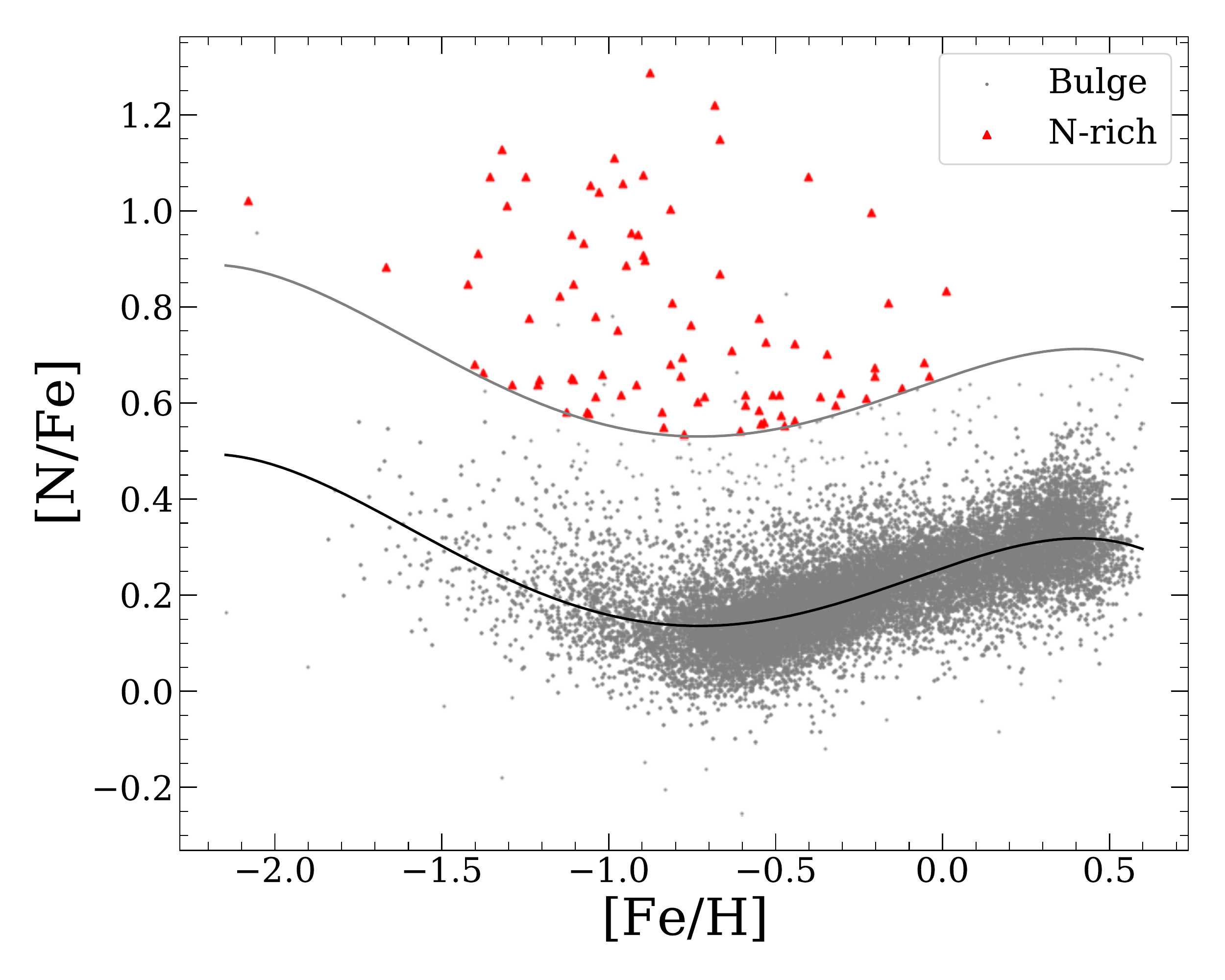}
    \caption{Distribution of sample stars in [N/Fe]-[Fe/H] plane. The small grey dots show the bulge population as selected in Section \ref{sample}. The red triangles indicate the N-rich stars, defined as stars which deviate from the $4^{th}$ order polynomial fit (black line) by more than $5.5\sigma$ and have [C/Fe] $<0.15$.}
    \label{nrich}
\end{figure}
\begin{figure*}
	\includegraphics[width=\textwidth]{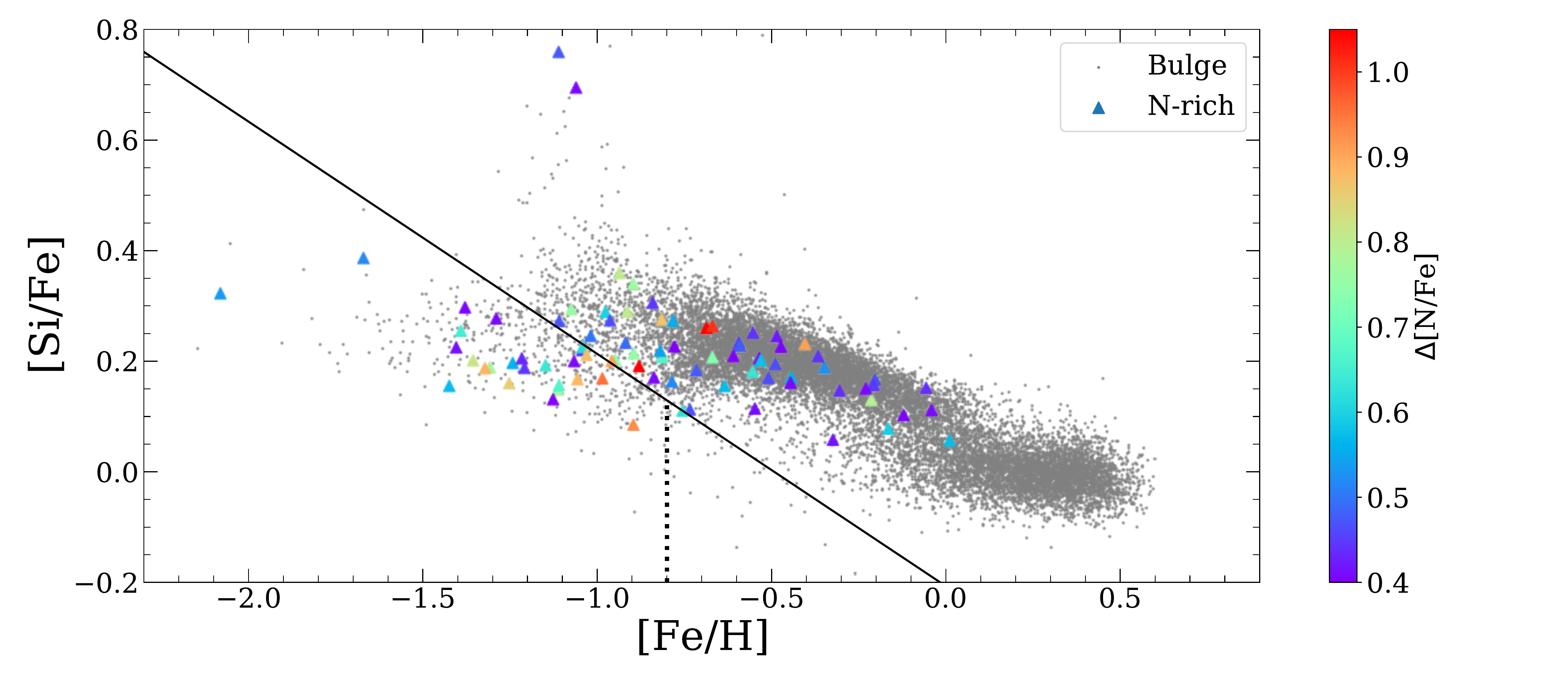}
    \caption{Distribution of sample stars in [Si/Fe]-[Fe/H] plane. The grey points show the distribution of the bulge stars, and the triangles, coloured according to the [N/Fe] residuals of the polynomial fit shown in Figure~\ref{nrich}, show the N-rich stars distribution in this plane. The solid black line is the cut made to separate the accreted stars from the \emph{in situ} stars shown in \citet{Mackereth2019}, adjusted to account for the metallicity gradient of disk populations between the outer and inner halo. The cut of [Fe/H] $<-0.8$ to remove disk contaminants is shown as the vertical dotted line.
    }
    \label{si}
\end{figure*}

\section{Results}
\label{result}
In this section, we discuss how our sample of accreted and \emph{in situ} populations are selected, employing methods used in \citet{Mackereth2019}. We then discuss how these populations differ from each other in orbital space, and show the similarities of the N-rich stars to GC members in chemical space.

\subsection{Selecting accreted and \emph{in situ} stars} 
\label{accselect}
In order to split our sample into accreted and {\it in situ} groups, we study the distribution of stars in the $\alpha$-Fe plane. \citet{Mackereth2019} achieved that by examining the distribution of their sample in the Mg-Fe plane, whereas \citet{hortaigs2020} focused on the distribution in the [Mg/Mn] vs [Al/Fe] plane. We cannot proceed in the same way, because the abundances of Al and Mg are affected by the multiple populations phenomenon in GCs (\citealp[e.g., ][]{Bastian-Lardo2018, Meszaros2015, Meszaros2020}), so that the positions of N-rich stars in chemical planes involving those elements cannot be interpreted in the same way as those of normal stars.  Therefore, we use Si as the tracer of $\alpha$-element abundances, because this element does not present substantial star-to-star variations in Galactic GCs.  

The data in Figure \ref{si} show that the N-rich star population occupies the same locus in the Si-Fe plane as the overall bulge field population. 
Following \citet{Mackereth2019}, we split the sample between accreted and {\it in situ} populations. 
To determine where the dividing line is drawn in the [Si/Fe] vs [Fe/H] plane, we proceed as follows: 
1) Following \citet{Mackereth2019}, we choose a slope that approximately matches the mean slope of the high- and low-$\alpha$ populations, slightly adjusting it to minimise the contamination of the accreted populations by low-$\alpha$ disk stars;
2) We calculate the distance in [Fe/H] between the dividing line and the mean value of the low-Mg disk population and adjust the zero-point so that the distance is the same in the [Si/Fe] vs [Fe/H] plane;
3) We further shift the zero-point by +0.2 dex in [Fe/H], to account for the disk metallicity gradient (\citealp[e.g., ][]{Hayden2015}).  The resulting linear relation is given by:
\begin{equation}
{\rm [Si/Fe]} \,=\, -0.42\,{\rm ([Fe/H]}\,+\, 0.016\,{\rm )} \,+\, 0.2
\label{line}
\end{equation}
Because this relation may be considered somewhat arbitrary, we estimate how a $\pm$0.1 dex zero-point variation impacts our results (see discussion in Section~\ref{kine})

We make a further cut in metallicity to the accreted population of [Fe/H] $<-0.8$, to minimise contamination from disk stars. This latter cut removes 38 bulge stars from our accreted population, bringing the total number of bulge stars down to 14,410.
We henceforth refer to stars below (above) and to the left (right) of the dividing line as "accreted" ({\it in situ}) populations.
The resulting accreted and \emph{in situ} general bulge samples comprise 428 and 13,982 stars, respectively, with 25 N-rich stars being located in the accreted locus, and 58 located in the {\it in situ} region. Thus, we conclude that roughly $\sim$30\% of the N-rich stars in the inner Galaxy have an accreted origin.  We emphasise here that stars in each sub-sample are found across the entire inner Galaxy.

\begin{figure}
	\includegraphics[width=\columnwidth]{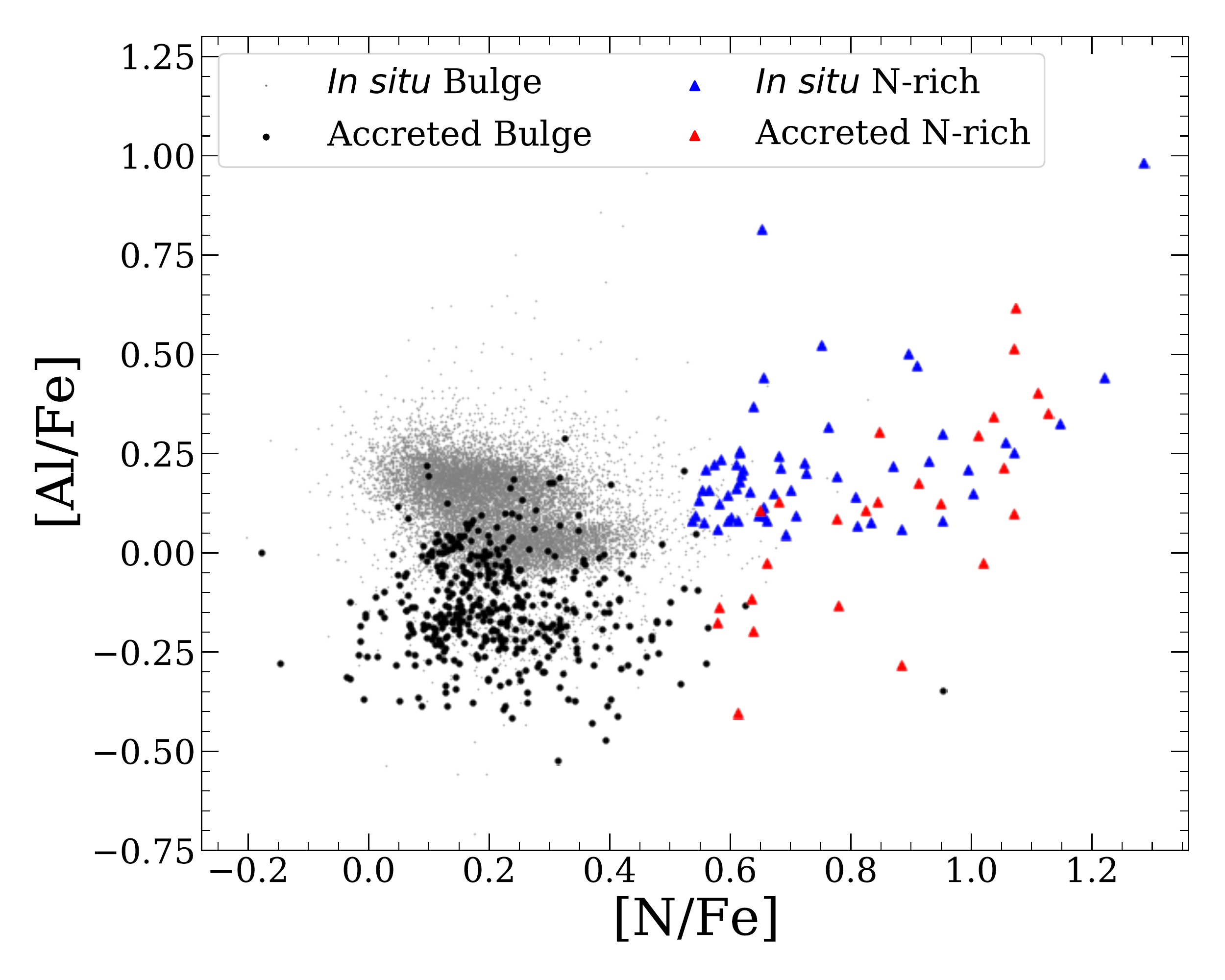}
    \caption{Distribution of \emph{in situ} bulge (small grey dots), accreted bulge (black dots), \emph{in situ} N-rich (blue triangles) and accreted N-rich (red triangles) stars in [Al/Fe]-[N/Fe] plane. The N-rich stars show a correlation between [N/Fe] and [Al/Fe], which is also observed in SG GC stars. However, the correlations are slightly different between the accreted and {\it in situ} populations. The accreted bulge stars are seen to occupy a lower locus in [Al/Fe] than the {\it in situ} by $\sim0.2$ dex, which is consistent with a dwarf galaxy origin.
    }
    \label{aln}
\end{figure}

Figure \ref{aln} shows where these sub-samples lie in the [Al/Fe]-[N/Fe] plane. By placing stars in this plane, former GC members can be identified as those that follow a positive correlation between those two abundance ratios.
When displaying our sample on this plane, we can see that the accreted and \emph{in situ} bulge populations occupy slightly different loci. 
While N and Al abundances of N-rich stars are correlated in both {\it in situ} and accreted sub-samples, the correlations in each sub-sample are slightly different.  The [Al/Fe] ratios of N-normal stars in the accreted sample, save for a handful of outliers, are lower than those in their {\it in situ} counterparts, on average by $\sim$ 0.2 dex.
This result validates our definition of accreted vs {\it in situ} populations, since the accreted stars with first-generation-like abundance patterns (i.e., those not affected by multiple population effects) are consistent with a dwarf galaxy origin (\citealp[e.g.,][]{Mackereth2019, Helmi2020, Das2020, hortaigs2020}). 

We identify a group of Si-rich stars, with [Si/Fe]~$\simgreater$~+0.5 in the metallicity range -1.3<[Fe/H]<-0.9.  They are  similar to those spotted by \citet{masseron2019} within the MW GCs M92, M15 and M13.  Those authors showed that, in the most metal-poor GCs, M92 and M15, Si-rich stars are characterised by very low [Mg/Fe], whereas stars in M13 had normal [Mg/Fe].  The Si-rich stars in our sample have normal [Mg/Fe], resembling those \citet{masseron2019} identified in M13.  We ascribe a GC origin to these field Si-rich stars and discuss their kinematic properties in Section \ref{kine}.

\begin{figure*}
	\includegraphics[scale=0.35]{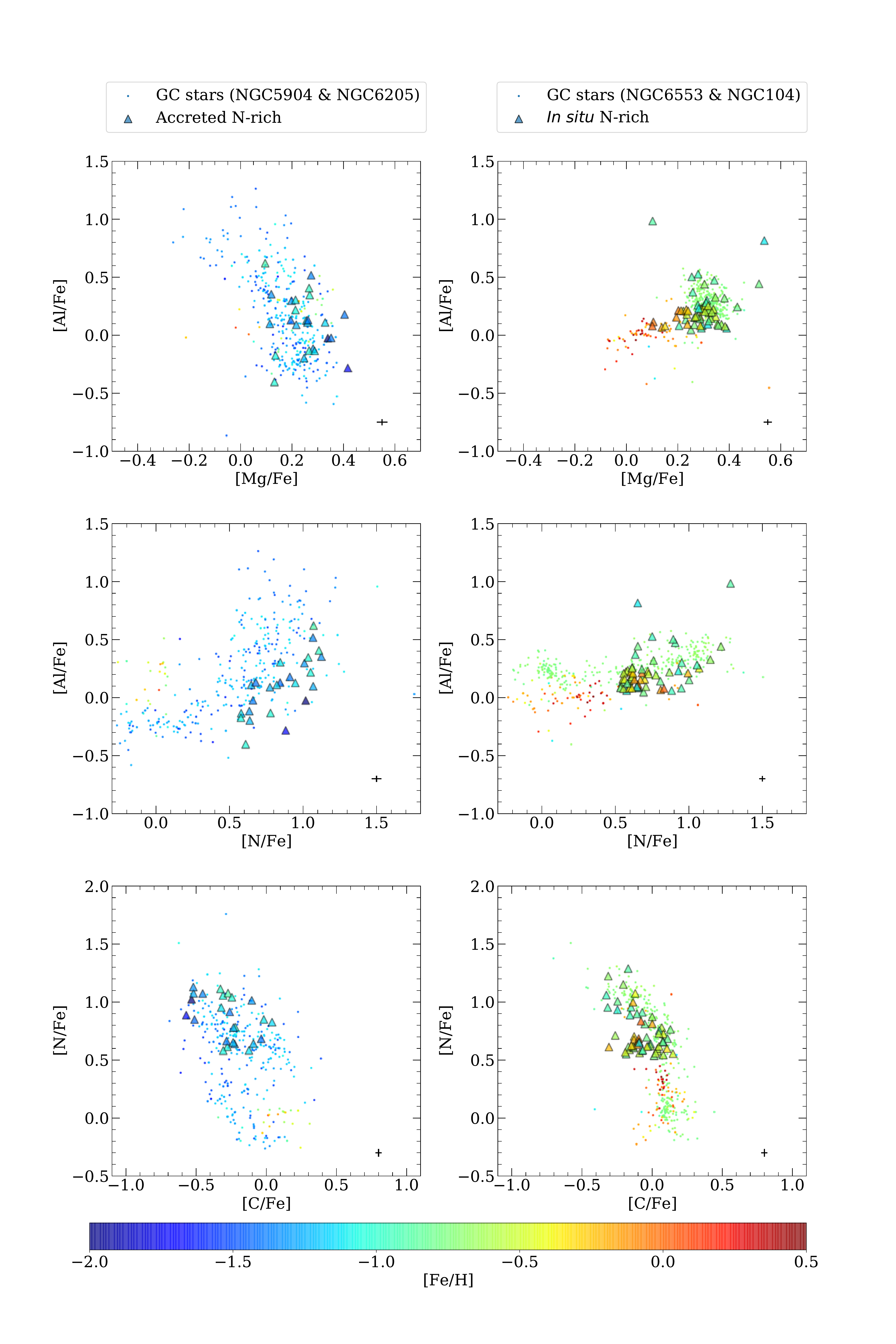}
    \caption{Coloured dots and triangles indicate GC stars \citep{HortaMNRAS2020} and N-rich stars (see Section \ref{sample}), respectively, colour coded by their [Fe/H] abundance. The graphs on the left show the accreted N-rich stars plotted on top of stars in NGC5904 and NGC6205, and the right graphs show the \emph{in situ} N-rich stars plotted on top of stars in NGC6553 and NGC104, both using the same metallicity colour scale.  Each plot shows the mean errorbar for the N-rich stars in the bottom right corner. 
    The $1^{st}$ row shows these stars in the [Al/Fe]-[Mg/Fe] plane to show the Al-Mg anti-correlation in GCs. 
    The $2^{nd}$ row shows the distribution in the [Al/Fe]-[N/Fe] plane to show the Al-N correlation in GCs. 
    The $3^{rd}$ row shows the distribution in the [N/Fe]-[C/Fe] plane to show the N-C anti-correlation in GCs.
    Each plot shows that our sample of N-rich stars lies on the same locus as SG GC members, supporting the idea they have possible GC origin.
    }
    \label{gcplot}
\end{figure*}

\subsection{Comparison with GCs}
\label{gc}
To confirm the association of the field N-rich stars with GCs, we overplot our sample of N-rich stars on data for GC members from \citet{HortaMNRAS2020} in three different chemical planes.  We show the correlations of GC stars in Mg-Al, Al-N and N-C space. In each panel the N-rich stars lie on the same locus as SG GC stars, which supports our assumption that they are, in fact, former GC members.  For clarity, the accreted and \emph{in situ} populations are plotted on different panels of Figure \ref{gcplot} because they span different metallicity regimes. Abundances of field stars in each set of panels are compared with those of members of GCs whose mean chemical compositions locate them in the accreted and {\it in situ} loci of the Si-Fe plane.
For the comparison with accreted N-rich stars we select NGC5904 (258 stars, <[Fe/H]> = -1.14, <[Si/Fe]> = 0.18) and NGC6205 (119 stars, <[Fe/H]> = -1.44, <[Si/Fe]> = 0.19), and for the {\it in situ} N-rich stars we select NGC6553 (52 stars, <[Fe/H]> = -0.04, <[Si/Fe]> = 0.06) and NGC104 (333 stars, <[Fe/H]> = -0.67, <[Si/Fe]> = 0.21)

On the plots in the first row, the anti-correlation between Al and Mg appears to differ sunstantially between the metal-poor and metal-rich sub-samples of GCs.  The metal-rich GC sub-sample shows a smaller scatter in both [Al/Fe] and [Mg/Fe] than those shown by the metal-poor sub-sample.  Therefore, while the anti-correlation is easily visible in the metal-poor sample, it is not evident in the metal-rich sample. 
This is similarly shown in the Al-N plots. Where, though the correlation can be seen in the metal-rich GCs, it is more easily identified in the metal-poor GC sub-sample.
For a more detailed discussion, see \citet{Meszaros2015} and \citet{Nataf2019}.

In a recent paper, \citet{Trincado2019} claim that N-rich stars must have [Al/Fe] > +0.5 to be considered SG GC members. Application of that criterion would remove large numbers of N-rich stars from our sample.  However, we argue that our sample of field N-rich are indeed akin to SG GC members for the following reason: the bottom panels of Figure \ref{gcplot} show a clear bimodality in the [N/Fe]-[C/Fe] plane, with the SG GC stars located and higher [N/Fe] above their FG GC counterparts. The dividing line between the two populations is located roughly at [N/Fe] = +0.5 for [C/Fe] = --0.5, and gently decreasing [N/Fe] for increasing [C/Fe]. This bimodality is also present in both the [Al/Fe]-[Mg/Fe] and [Al/Fe]-[N/Fe], showing that there are SG GC stars with [Al/Fe]<0.5 all the way to below solar. In fact, application of an [Al/Fe] > +0.5 cut would remove a large fraction of the SG stars in GCs themselves, particularly in the low metallicity regime (left panels of Figure \ref{gcplot}). It is also well known that, although SG GCs typically present enhancements in N, Al and Na \citep{Bastian-Lardo2018}, not all stars in GCs that are enhanced in N are also enhanced in Al.  Indeed, as mentioned above, the Al-Mg anti-correlation is dependent on metallicity, being substantially weaker in metal-rich GCs (\citealp[e.g.,][]{Meszaros2015, Nataf2019, Meszaros2020}), and mass \citep{massari2017}.

\begin{figure}
	\includegraphics[width=\columnwidth]{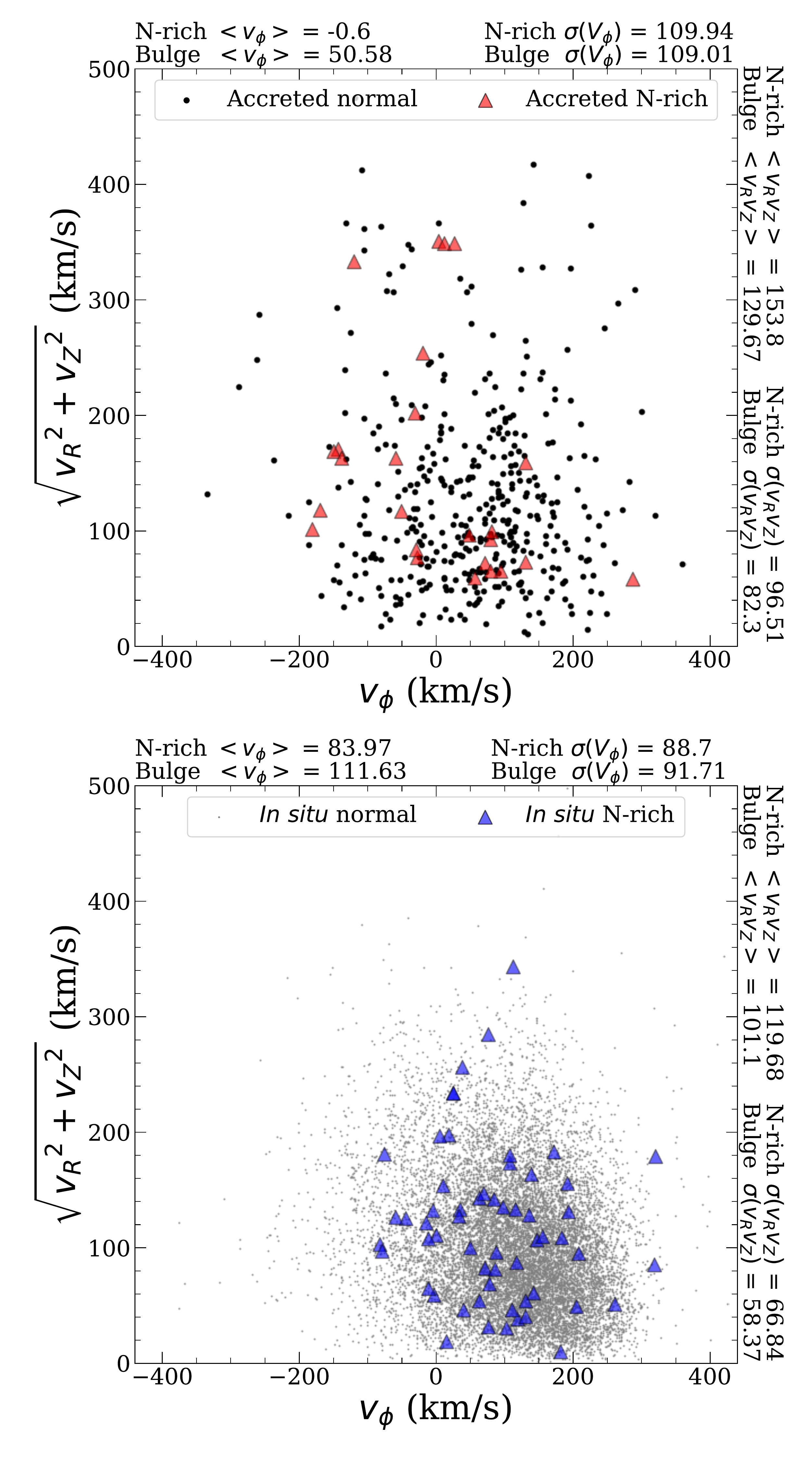}
    \caption{Distribution in $\sqrt{{v_R}^2 + {v_Z}^2}$ vs. $v_\phi$ of accreted and \emph{in situ} stars on the top and bottom panel respectively. \emph{Top Panel:} Accreted N-rich stars (red triangles) and the accreted bulge stars (black dots). \emph{Bottom Panel:} \emph{in situ} N-rich stars (blue triangles) and \emph{in situ} bulge stars (grey dots). Anything with $v_\phi$ > 0 has a prograde orbit, similar to that of the disk, and anything with $v_\phi$ < 0 has a retrograde orbit. We also show on these plots the mean and standard deviation of the sub-samples for each axis.
    }
    \label{vel}
\end{figure}

\subsection{Kinematic properties}
\label{kine}
In this sub-section we check whether our definition of accreted and \emph{in situ} stars, which is based solely on chemistry, maps into distinct properties in kinematic space. 
To do this, we make comparisons between the distributions of our samples in a kinematic diagram, which is used to distinguish components of the Galaxy on the basis of their kinematic signatures (\citealp[e.g.,][]{Venn2004, Bonaca2017, Helmi2018, Koppelman2019}). The x-axis of the kinematic diagram is the tangential velocity, $v_\phi$, while the y-axis is the quadrature sum of the radial and vertical velocities, $\sqrt{{v_R}^2 + {v_Z}^2}$.

The accreted and \emph{in situ} populations are displayed on the kinematic diagram separately on the upper and lower panels of Figure \ref{vel}, respectively.  Since the velocities are in Galactocentric coordinates, this places the origin of the coordinate system at the Galactic Centre, therefore the velocity of the Sun is at ${\rm v}_{LSR} \sim 220$~km/s.  In both panels normal stars are displayed as black/gray dots and N-rich stars as coloured triangles.  Visual examination of these plots suggests the following interesting trend:  Accreted stars, both normal and N-rich, have on average more retrograde orbits (${\rm v}_\phi < 0$) than their {\it in situ} counterparts, whose orbits are predominantly prograde. This is clearly shown by the difference in the $v_\phi$ distribution of the {\it in situ} and accreted samples of N-rich stars, with the mean of the latter being $\sim$ 80 km/s lower than that of the former.

\begin{figure}
	\includegraphics[width=\columnwidth]{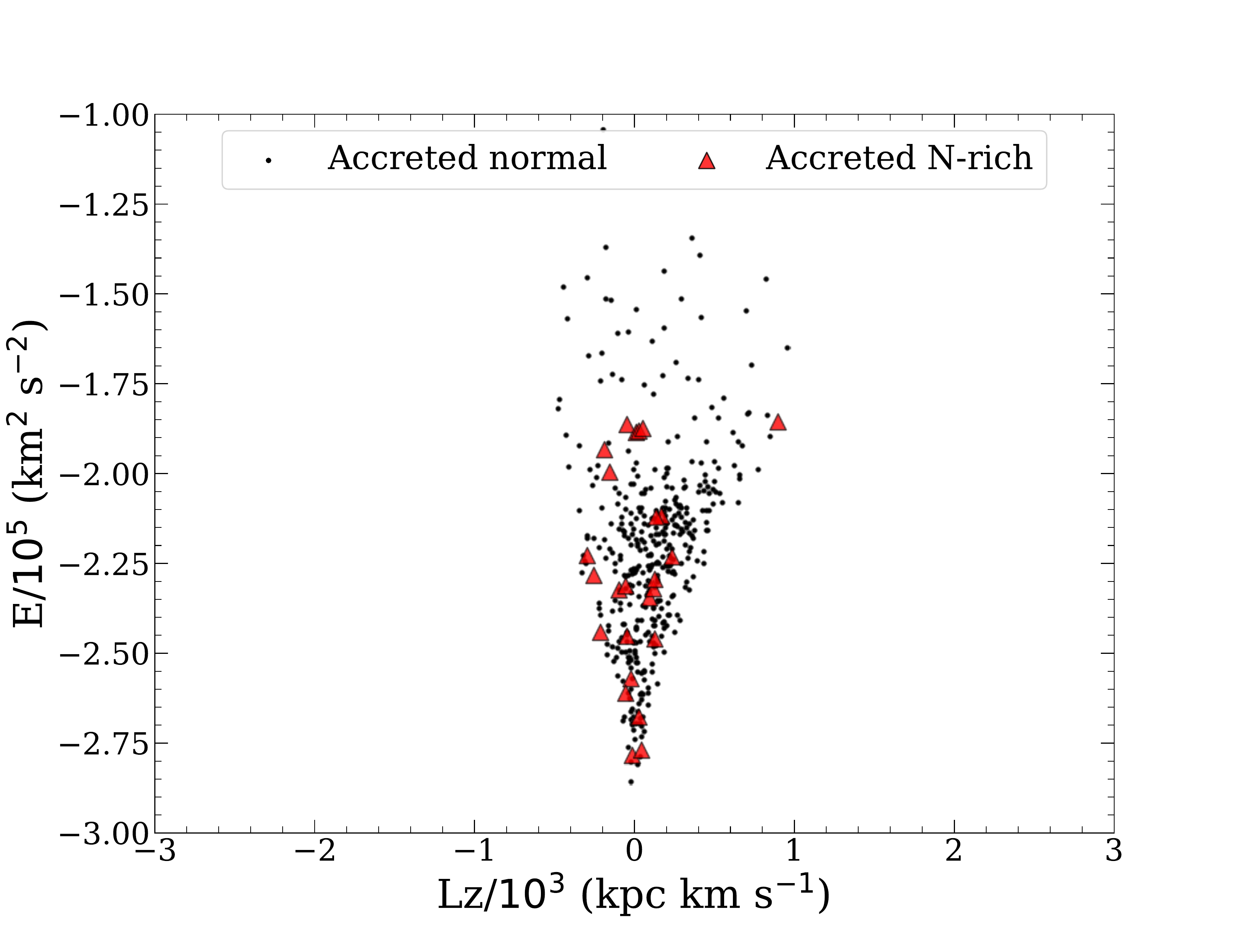}
    \caption{Accreted N-normal (black dots) shows a population of stars in the same locus as GE/S stars at high energies, whilst the accreted N-rich stars (red triangles) occupy the low energy region similar to Heracles.
    }
    \label{elz}
\end{figure}

The above visual impressions must be confirmed by a quantitative statistical evaluation.  The Kolmogorov-Smirnov (KS) statistic is a nonparametric test used to assess the similarity between two samples.
We use the python package \texttt{ndtest}\textit{}\footnote{\url{https://github.com/syrte/ndtest}} to make 2D comparisons between the distributions in $v_\phi$ and $\sqrt{{v_R}^2 + {v_Z}^2}$ of the following sub-samples, as shown in Table \ref{kstest}: accreted N-rich vs. accreted normal, \emph{in situ} N-rich vs. \emph{in situ} normal, accreted N-rich vs. \emph{in situ} N-rich and, accreted normal vs. \emph{in situ} normal.
The KS tests result in a rejection of the null hypothesis, with $p$-value < 0.1 for all four comparisons. The clear kinematic distinction between the accreted and {\it in situ} populations confirms our chemical selection of these groups. 
We also note the difference between accreted N-rich vs. accreted normal sub-samples. This result can be understood by examination of Figure \ref{elz}.  
In that plot it can be seen that the accreted normal stars show a clump of slightly prograde stars around $E/10^5 \sim -2.2$~km$^2$s$^{-2}$, without a clear counterpart in the N-rich accreted group.  We suspect that this prograde population is likely due to contamination from the disk.  In addition, the accreted normal population hosts a number of stars forming a cloud with $E/10^5 \simgreater -1.85$~km$^2$s$^{-2}$, where no N-rich stars can be found.  That is the locus occupied by stars belonging to the GE/S system, as well as other possible accretion events \citep{Ibata1994, Helmi1999, Ibata2016, Belokurov2018, haywood2018, Helmi2018, Mackereth2019, hortaigs2020}. Conversely, most of the N-rich stars occupy the same locus as Heracles identified recently by \citet{hortaigs2020}, with a couple of stars displaying kinematics suggestive of disk-like orbits.

\begin{table}
    \centering
    \resizebox{8cm}{!}{
    \begin{tabular}{l|r}
    \hline
    {\bf Comparison} & {\bf $p$-value} \\
    \hline
    Accreted N-rich vs. Accreted normal & 0.024 \\
    \emph{In situ} N-rich vs. \emph{In situ} normal & 0.028 \\
    Accreted normal vs. \emph{In situ} normal & $<0.001$ \\
    Accreted N-rich vs. \emph{In situ} N-rich & 0.009 \\
    \hline 
    \hline
    {\bf Comparison ( [Fe/H] < --0.8 )} & {\bf $p$-value} \\
    \hline
    Accreted N-rich vs. Accreted normal & 0.024 \\
    \emph{In situ} N-rich vs. \emph{In situ} normal & 0.294 \\
    Accreted normal vs. \emph{In situ} normal & 0.038 \\
    Accreted N-rich vs. \emph{In situ} N-rich & 0.027 \\
    \hline
    \hline
    {\bf Comparison ( Zero-point +0.1 dex )} & {\bf $p$-value} \\
    \hline
    \hline
    Accreted N-rich vs. Accreted normal & 0.066 \\
    \emph{In situ} N-rich vs. \emph{In situ} normal & 0.062 \\
    Accreted normal vs. \emph{In situ} normal & $<0.001$ \\
    Accreted N-rich vs. \emph{In situ} N-rich & 0.040 \\
    \hline
    {\bf Comparison ( Zero-point --0.1 dex )} & {\bf $p$-value} \\
    \hline
    Accreted N-rich vs. Accreted normal & 0.172 \\
    \emph{In situ} N-rich vs. \emph{In situ} normal & 0.006 \\
    Accreted normal vs. \emph{In situ} normal & $<0.001$ \\
    Accreted N-rich vs. \emph{In situ} N-rich & 0.223 \\
    \hline
    \end{tabular}{}
    }
    \caption{Results obtained from performing a 2D KS test between the different sub-samples shown in Figure \ref{vel}. \emph{First Panel:} $p$-values for the comparisons between sub-samples as defined in Section \ref{accselect}. \emph{Second Panel:} $p$-values for the comparisons between the sub-samples with [Fe/H] $<-0.8$. 
    \emph{Third \& Fourth Panel:} Result when shifting the zero-point of the dividing line, Equation~\ref{line}, by $\pm$0.1~dex.
    Setting a threshold for the $p$-value of 0.1. So, a $p$-value < 0.1 results in a rejection of the null hypothesis, whereas a $p$-value > 0.1 means the null hypothesis cannot be rejected.}
    \label{kstest}
\end{table}

Interestingly, the KS test rejects the null hypothesis for similarity between the \emph{in situ} N-rich vs. \emph{in situ} normal sub-samples. We suggest that the difference between these two sub-samples is due to the presence of disk stars within 4 kpc of the Galactic centre. This is further discussed in Section \ref{freq} 

The stars in the accreted sample have on average lower metallicities than their {\it in situ} counterparts. Thus, the differences encountered could be due to the dependence of kinematics on the metallicity of stellar populations. To test that hypothesis, we redo the KS tests to assess the similarity between the accreted and in situ sub-samples, this time limiting the comparison to stars with [Fe/H] $<-0.8$. The results from this comparison are shown in Table \ref{kstest}. The difference between the N-rich and normal accreted populations remain unchanged since they were already restricted to [Fe/H] $<-0.8$. We do, however, see a big change in the comparison between the {\it in situ} populations, where the $p$-value = 0.294 tells us that the null hypothesis cannot be rejected. This is due to the removal of high metallicity disk stars from our sample of {\it in situ} normal stars.  Regarding the comparison between accreted and {\it in situ} populations, for both the N-rich and normal samples, the null hypothesis is rejected even when the comparison is limited to metal-poor sub-samples.  In short,  accreted and {\it in situ} samples are kinematically different populations even when only metal-poor stars are considered.

We checked whether our results are sensitive to the definition of the line separating accreted and {\it in situ} populations in Figure~\ref{si}.  For that purpose, we shifted the zero-point of the relation given by the Equation~\ref{line} by $\pm$0.1~dex in [Fe/H], the results for which are shown in the bottom two panels of Table~\ref{kstest}.  When increasing the zero-point by +0.1~dex, our results are unchanged.  However, when shifting the relation towards lower [Fe/H], the KS tests become consistent with the null hypothesis for two of the sub-sample comparisons: {\it (i)} accreted normal vs. accreted N-rich stars.  This result is due to the removal of a small number of retrograde N-rich stars and the reduction in the contribution of prograde normal stars (which we conjectured in Section~\ref{kine} to be due to disk contamination); {\it (ii)} accreted N-rich vs. {\it in situ} N-rich stars.  This happens because the above mentioned retrograde N-rich stars that are moved from the accreted to the {\it in situ} sub-sample, make the two groups more similar kinematically.  Since this exercise leads to a reduction of the size of the N-rich accreted population, we deem these result of little statistical significance. The matter will have to be revisited on the basis of larger samples.

Again, we check the dependence of kinematics on metallicity by limiting the comparison to stars with [Fe/H] $<-0.8$, as done above, after shifting the relation by $\pm$0.1~dex.  The results for this are not shown since the only change we find is when comparing {\it in situ} N-rich and {\it in situ} normal sub-samples. In both cases, when moving the zero-point towards higher of lower [Fe/H], the null hypothesis cannot be rejected when comparing these two sub-samples.  Also in this case the statistical significance of the results is small due to the reduced sample sizes.

Finally, we examine the kinematic properties of Si-rich stars mentioned in Section \ref{accselect} separately.  When comparing their properties to those of N-rich and N-normal, the KS tests only yielded a statistically significant difference with the accreted N-rich, p-value=0.022. This suggests that this population is likely to result from the dissolution of {\it in situ} GCs and in the remainder of this analysis they will be treated as such.

In summary, the results above show that the chemistry-based definition of accreted and {\it in situ} sub-samples maps into distinct kinematic properties. Both N-rich and N-normal {\it in situ} samples with $R_{\rm GC}<4~{\rm kpc}$ show more disk-like orbits than their accreted counterparts, according to expectations. In \citet{hortaigs2020} we showed that there is an important contamination of the chemically defined accreted samples by {\it in situ} stars.  However, the differences persist even when controlling for the dependence of kinematics on metallicity, which argues in favour of our interpretation of the origin of the accreted N-rich sample. 

\subsection{N-rich stars frequency in accreted and  \emph{in situ} samples}
\label{freq}

\begin{figure}
	\includegraphics[width=\columnwidth]{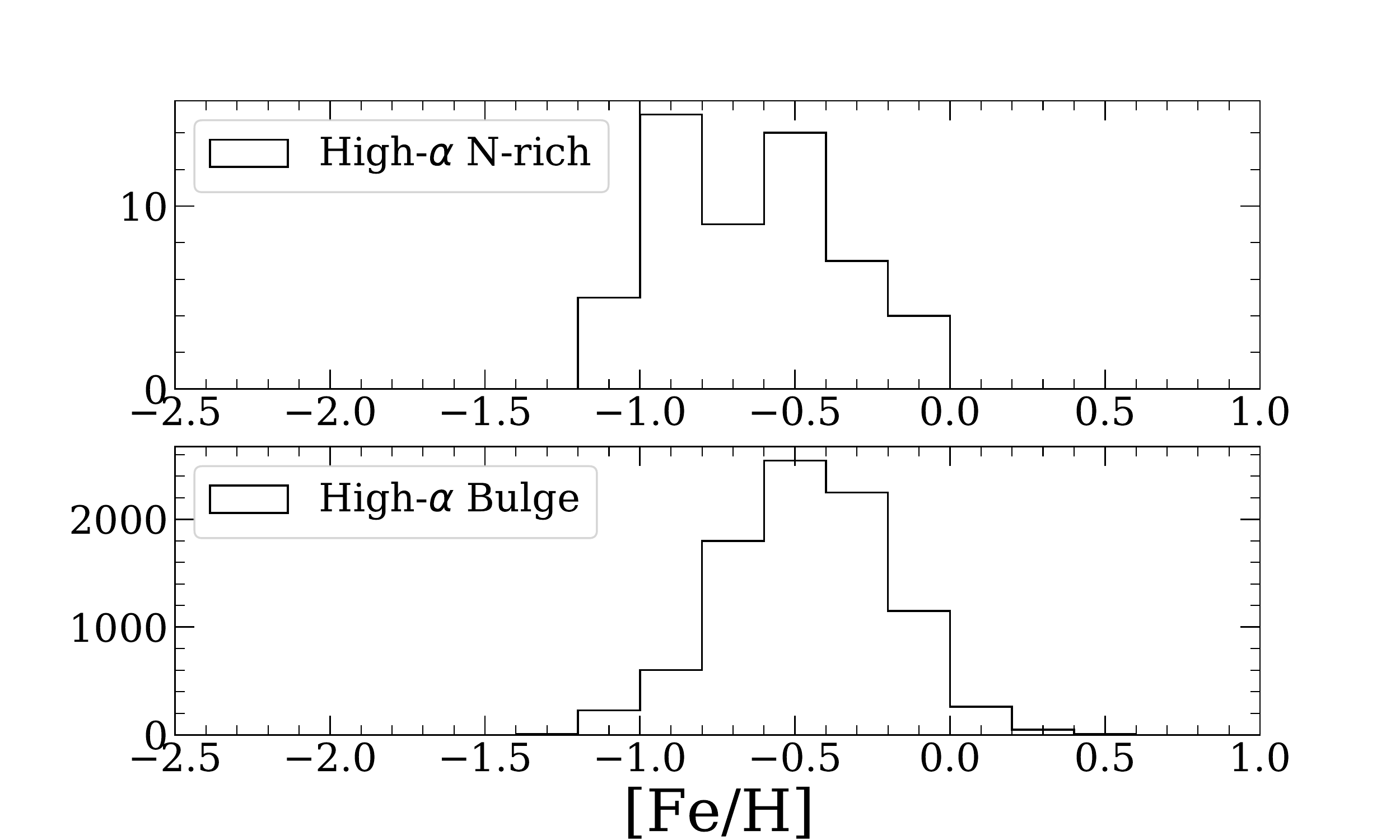}
    \caption{Metallicity distribution functions (MDFs) for the \emph{in situ} high-$\alpha$ N-rich stars (top panel) and \emph{in situ} high-$\alpha$ normal field (bottom panel). The MDFs of the two populations are not very different.  The N-rich MDF peaks at a slightly lower [Fe/H], but is substantially broader, overlapping the high metallicity end of the N-normal MDF.
    }
    \label{mdf}
\end{figure}

An important clue to the origin of N-rich stars is their frequency, $f_{Nr}$, defined as the ratio between the number of such stars and the total field population (\citealp[e.g.,][]{martell2016, Schiavon2017, Koch2019, density2020}).  We measured this frequency in both the accreted and {\it in situ} sub-samples, and henceforth express it in terms of percentages.  In the accreted group we find $f_{Nr}=5.84\pm 1.28$\%, whereas for the {\it in situ} group, the measured frequency is an order of magnitude lower, $f_{Nr}=0.41\pm 0.05$\%, $f_{Nr}=0.60\pm 0.08$\% if only high-$\alpha$ stars are considered.
If we account for the Si-rich stars identified in Section \ref{accselect}, ascribing them to an {\it in situ} GC origin based on their kinematics, the frequency of the {\it in situ} group increases to  $f_{Nr}=0.58\pm 0.06$\%, $f_{Nr}=0.86\pm 0.10$\% if only high-$\alpha$ stars are considered. Thus, consideration of Si-rich stars does not alter our finding of a large difference between accreted and {\it in situ} N-rich stars.

This difference cannot be easily understood.  According to the prevailing scenario for GC formation and destruction, \citep[][]{kruijssen2014, kruijssen2015,pfeffer2019} Galactic GCs originate from two different channels. The {\it ex situ}, or accreted, channel would consist of GCs that were accreted to the Galaxy along with their host galaxies.  Those accretion episodes occurred predominantly, though not exclusively, in the early stages of the Milky Way assembly, as suggested by various lines of evidence \citep[e.g.,][]{Deason2014,Mackereth2018,Mackereth2019,pfeffer2019,Schiavon2020,hughes2020}. Conversely, the {\it in situ} population would be comprised of GCs that were formed in the turbulent disk of the Milky Way at $z \sim 2-3$. According to this scenario, {\it in situ} GCs would have been destroyed very efficiently by tidal interaction with giant molecular clouds in the early disk \cite[the so-called ``cruel cradle effect'', see][]{cce2012}, whereas destruction of accreted GCs via tidal stripping and evaporation was less efficient, having happened on a much longer timescale.  Given these predictions, we would naively expect the frequency of {\it in situ} N-rich stars to be higher, not lower than that of the accreted population.

One possible way out of this conundrum is to invoke that the ratio between integrated star formation in the form of GCs over total was lower in the {\it in situ} than in the accreted population.  This could be achieved, for instance, if the {\it in situ} population underwent a longer star formation episode than that leading up to the formation of the accreted population.  If  {\it in situ} star formation was extended further in time, after the cessation of the main episode of GC formation/destruction, a low {\it in situ} $f_{Nr}$ could possibly be accommodated.  In such a situation, however, one would expect the metallicity distribution function (MDF) of the {\it in situ} normal population to have more power towards higher metallicities than that of the {\it in situ} N-rich population.  

This qualitative prediction does not seem to be supported by the metallicity distribution functions (MDFs) of the N-rich and N-normal bulge {\it in situ} samples, shown in Figure~\ref{mdf}.  For simplicity, we limit our comparison to high-$\alpha$ N-rich and normal field stars, as those are understood to have undergone a coherent chemical evolution path that is independent of the low-$\alpha$ disk population \cite[][]{Mackereth2018}.  In that figure, one can see that the MDFs of the high-$\alpha$ N-rich and N-normal {\it in situ} samples are not very different.  The MDF of the high-$\alpha$ N-rich population peaks at [Fe/H]$\sim$--0.8, whereas that of the high-$\alpha$ N-normal population peaks at a slightly higher metallicity, around [Fe/H]$\sim$--0.5.  That difference in the mode of the two MDFs is slightly offset by the fact that the N-rich population has a broader MDF, with FWHM$\sim$0.9~dex, whereas that of the {\it in situ} population has FWHM$\sim$0.6~dex. Assuming the N-rich MDF reproduces that of the parent GCs, one would thus conclude that the star formation history associated with the N-normal population did not extend in time much further past the period of GC formation and destruction.  This result is additionally corroborated by recent evidence for a very fast overall formation of both the high- and low-$\alpha$ stellar populations in the inner Galaxy, which is attested by their predominantly old ages (\citealp[e.g.,][]{Hasselquist2020}).  

This result prompts interesting considerations on the origin of the accreted N-rich stars currently inhabiting the inner Galaxy.  The frequency of metal-poor N-rich stars as a function of Galactocentric distance has been shown by \citet{density2020} to undergo a steep decrease towards growing $R_{\rm GC}$ \citep[see also][]{martell2016,Koch2019}.  At $R_{\rm GC}\sim15~{\rm kpc}$, \citet{density2020} found $f_{Nr}\sim3^{+1}_{-0.8}$\%, which is considerably lower than the ratio we find for the accreted population. Since the population of N-rich stars in our sample at low metallicity is dominated by accreted stars, this result leads to the conclusion that GC destruction associated with satellite mergers must have been very efficient in the early stages of the Galaxy's formation.  Indeed it has been shown by \citet{Pfeffer2020} that GCs associated with the earliest accretion events ended up in strongly bound orbits, driven by dynamical friction.  That is the case for Heralces \citep{hortaigs2020}, a $\sim 5\times10^8$~M$_\odot$ satellite that likely merged with the MW over 10 Gyr ago \citep[see also][]{Kruijssen2020}.  Given the coincidence between the positions of our bulge N-rich stars in integrals of motion space and those of Heracles stars (Figure~\ref{elz}), we speculate that the bulge N-rich population is partly made of members of GCs that were originally associated with Heracles, and were mostly destroyed during the accretion event.  It is also possible that those accreted N-rich stars were already in the field of Heracles, before they were accreted to the MW, however there currently is no evidence for the presence of N-rich stars in the fields of dwarf satellites of the MW.

\cite{hughes2020} used the E-MOSAICS simulations \citep{pfeffer2019} to calculate the contribution of destroyed GCs to field populations in the bulges of MW-like galaxies, comparing the predictions with the measurements by \cite{Schiavon2017}.  They show that, for most of the MW-like galaxies in their simulated volume, the prediction for $f_{Nr}$ of the {\it metal-poor} stellar population is lower than the observations by factors of $\sim$2--30 (bottom panel of their Figure 4).  However, for a few simulated galaxies the predicted $f_{Nr}$ are in good agreement with the observations. Like the MW, the disk populations of those galaxies are characterised by a bimodal distribution in the $\alpha$-Fe plane, which is a distinctive feature of the MW disk populations \citep[e.g.,][]{Hayden2015,Mackereth2017}.  \citet{Mackereth2018} showed that this feature is associated with an atypical accretion history, characterised by intense merging in early times and relative calm since $z\sim1-1.5$.  It is noteworthy, however, that \cite{hughes2020} predictions for these few MW-like galaxies differ from our measurements with regards to the dependence of $f_{Nr}$ on position in the $\alpha$-Fe plane.  The high frequency of ex-GC stars in the field of simulated galaxies is predominantly due to high-$\alpha$ {\it in situ} GC formation and destruction, whereas our data show that the high $f_{Nr}$ in the MW bulge is due to the contribution by the dissolution of low-$\alpha$ {\it accreted} GCs.  This discrepancy would be alleviated if some of the stars in the accreted region in Figure~\ref{si} were in fact formed {\it in situ}, \citep[see Figure 2 of][]{hughes2020}, but it is not clear that accounting for such a contamination would completely eliminate the disagreement.

\section{Summary}
\label{sum}
The results presented in this paper make use of elemental abundances from APOGEE DR16 along with data from {\it Gaia} DR2 to study the chemical and kinematic properties of 146 N-rich stars located within the inner 4 kpc of the Galaxy.
Our conclusions can be summarised as follows:

\begin{itemize}
  \item We find that there are likely accreted and \emph{in situ} components to the N-rich population within 4 kpc of the Galactic centre, identified via chemistry by making a cut in [$\alpha$/Fe]-[Fe/H] space towards low metallicities (as shown in Figure \ref{si}) \citep[e.g.][]{hayes2018, Mackereth2019, Das2020}. By making this cut and removing stars without proper motions in {\it Gaia}, we select 428 and 13,982 bulge stars that lie in the accreted and {\it in situ} positions, respectively, with 25 N-rich stars being located in the accreted, and 58 located in the {\it in situ} locus.
  
  \item We show that our sample of N-rich stars occupies the same locus as so-called second-generation GC stars, supporting the idea that they are the by-products of GCs destruction/evaporation.  
  
  \item We find that there is a significant difference in the kinematic properties of chemically defined accreted and {\it in situ} populations. This shows that our chemistry-based distinction of these populations maps into differences in kinematic space. We also find that the accreted bulge field population includes stars which share orbital properties with the GE/S system, although no N-rich stars occupy that locus of orbital parameter space. The absence of N-rich stars associated with GE/S in the bulge is likely due to their low frequency, combined with the relatively small number of GE/S stars found in the bulge \citep[see][]{hortaigs2020}
  
  \item We find that the frequency of N-rich stars differs by an order of magnitude between the accreted ($f_{Nr}=5.84\pm 1.28$\%)  and {\it in situ} ($f_{Nr}=0.41\pm 0.05$\%) samples.  This result seems to be at odds with numerical simulations that predict a higher frequency of destroyed GCs among high-$\alpha$ {\it in situ} populations \citep{hughes2020}. 
  We speculate that the higher frequency of N-rich stars among accreted populations is due to early merger events, such as Heracles \citep{hortaigs2020}, which likely had their GCs destroyed very efficiently during the merger with the MW.  
  
  \item The identification of an accreted population of N-rich stars in the bulge  raises the question of whether the GCs from which they originate were destroyed in their host dwarf galaxies or during the merger. If the former hypothesis is correct, we would expect that N-rich stars would be present in the field of current Milky Way satellites. \citet{norris2017} did not find a Na-O anti-correlation, which is typical of GC stars, in Carina dwarf spheroidal field stars. However, their study is based on a sample of 63 stars, which is relatively small. Since the observed frequency of N-rich stars in the halo is $\sim3\%$ one would expect to find $\sim2$ N-rich stars in the sample of \citet{norris2017}. Such low numbers could easily be missed due to stochastic sampling.

\end{itemize}

\section*{Acknowledgements}

We thank all those professionals who have been working tirelessly during these difficult times so that people like ourselves can work safely from home.  The authors thank Nate Bastian and Meghan Hughes for helpful discussion. 
SSK acknowledges an STFC doctoral studentship.  The anonymous referee is thanked for an insightful review of the original manuscript.
JTM acknowledges support from the ERC Consolidator Grant funding scheme (project ASTEROCHRONOMETRY, \url{https://www.asterochronometry.eu}, G.A. n. 772293).
S.H. is supported by an NSF Astronomy and Astrophysics Postdoctoral Fellowship under award AST-1801940.
DAGH acknowledges support from the State Research Agency (AEI) of the  
Spanish Ministry of Science, Innovation and Universities (MCIU) and  
the European Regional Development Fund (FEDER) under grant AYA2017-88254-P.

Funding for the Sloan Digital Sky Survey IV has been provided by the Alfred P. Sloan Foundation, the U.S. Department of Energy Office of Science, and the Participating Institutions. SDSS acknowledges support and resources from the Center for High-Performance Computing at the University of Utah. The SDSS web site is www.sdss.org.
\\
SDSS is managed by the Astrophysical Research Consortium for the Participating Institutions of the SDSS Collaboration including the Brazilian Participation Group, the Carnegie Institution for Science, Carnegie Mellon University, the Chilean Participation Group, the French Participation Group, Harvard-Smithsonian Center for Astrophysics, Instituto de Astrof\'\i sica de Canarias, The Johns Hopkins University, Kavli Institute for the Physics and Mathematics of the Universe (IPMU) / University of Tokyo, the Korean Participation Group, Lawrence Berkeley National Laboratory, Leibniz Institut f\"ur Astrophysik Potsdam (AIP), Max-Planck-Institut f\"ur Astronomie (MPIA Heidelberg), Max-Planck-Institut f\"ur Astrophysik (MPA Garching), Max-Planck-Institut f\"ur Extraterrestrische Physik (MPE), National Astronomical Observatories of China, New Mexico State University, New York University, University of Notre Dame, Observat\'orio Nacional / MCTI, The Ohio State University, Pennsylvania State University, Shanghai Astronomical Observatory, United Kingdom Participation Group, Universidad Nacional Aut\'onoma de M\'exico, University of Arizona, University of Colorado Boulder, University of Oxford, University of Portsmouth, University of Utah, University of Virginia, University of Washington, University of Wisconsin, Vanderbilt University, and Yale University.

\section*{Data Availability}
Most of the data upon which this paper is based are publicly available as part of the 16th data release of the Sloan Digital Sky Survey (SDSS-IV) collaboration.  For part of the sample, the data are still proprietary and will be made publicly available as part of the 17th data release.  Once the latter data are publicly available they will be accessible via the usual channels. 

%%%%%%%%%%%%%%%%%%%%%%%%%%%%%%%%%%%%%%%%%%%%%%%%%%

%%%%%%%%%%%%%%%%%%%% REFERENCES %%%%%%%%%%%%%%%%%%

% The best way to enter references is to use BibTeX:

\bibliographystyle{mnras}
\bibliography{citations} % if your bibtex file is called example.bib

% Alternatively you could enter them by hand, like this:
% This method is tedious and prone to error if you have lots of references
%\begin{thebibliography}{99}
%\bibitem[\protect\citepauthoryear{Author}{2012}]{Author2012}
%Author A.~N., 2013, Journal of Improbable Astronomy, 1, 1
%\bibitem[\protect\citepauthoryear{Others}{2013}]{Others2013}
%Others S., 2012, Journal of Interesting Stuff, 17, 198
%\end{thebibliography}

%%%%%%%%%%%%%%%%%%%%%%%%%%%%%%%%%%%%%%%%%%%%%%%%%%

%%%%%%%%%%%%%%%%% APPENDICES %%%%%%%%%%%%%%%%%%%%%
\appendix
\section{Table of APOGEE IDs for N-rich stars}
Table \ref{ids} shows only the publicly available DR16 APOGEE ID's, RA and DEC for the N-rich stars selected in this paper.

%If you want to present additional material which would interrupt the flow of the main paper, it can be placed in an Appendix which appears after the list of references.

%%%%%%%%%%%%%%%%%%%%%%%%%%%%%%%%%%%%%%%%%%%%%%%%%%

% Don't change these lines
\bsp	% typesetting comment
\label{lastpage}
\begin{table*} 
 \centering
  \caption{N-rich stars identified in the inner Galaxy.}
  \begin{tabular}{cll}
  \hline
APOGEE\_ID  & RA & DEC \\ 
 \hline
  2M16051144-2330557 & 241.297673 & -23.515484 \\ 
  2M16180906-2442217 & 244.537768 & -24.706036 \\ 
  2M16304650-2949522 & 247.693763 & -29.831173 \\ 
  2M16314726-2945273 & 247.946932 & -29.757591 \\ 
  2M16333703-3028333 & 248.404329 & -30.475943 \\ 
  2M16335569-1344044 & 248.482062 & -13.734557 \\ 
  2M17024730-2210387 & 255.697092 & -22.177443 \\ 
  2M17271907-2718040 & 261.829481 & -27.301126 \\ 
  2M17281699-3024573 & 262.070794 & -30.415928 \\ 
  2M17285196-2013080 & 262.2165 & -20.218908 \\ 
  2M17293012-3006008 & 262.375515 & -30.100246 \\ 
  2M17293730-2725594 & 262.405434 & -27.433182 \\ 
  2M17303980-2330234 & 262.665839 & -23.506523 \\ 
  2M17305251-2651528 & 262.718823 & -26.864672 \\ 
  2M17305645-3030155 & 262.73523 & -30.504309 \\ 
  2M17325943-3034281 & 263.247636 & -30.57449 \\ 
  2M17330999-1034023 & 263.291625 & -10.567309 \\ 
  2M17333623-2548156 & 263.400967 & -25.804361 \\ 
  2M17334418-3033313 & 263.434107 & -30.558695 \\ 
  2M17334704-3034136 & 263.446029 & -30.570456 \\ 
  2M17335209-3011013 & 263.467059 & -30.183704  \\ 
  2M17340261-2616237 & 263.51091  & -26.273256 \\ 
  2M17343807-2557555 & 263.658637 & -25.965429 \\ 
  2M17350460-2856477 & 263.769185 & -28.946587 \\ 
  2M17354063-3339547 & 263.919305 & -33.665203 \\ 
  2M17404143-2714570 & 265.172631 & -27.249172 \\ 
  2M17494963-2318560 & 267.4568 & -23.315571 \\ 
  2M17504980-2255083 & 267.70754 & -22.91898 \\ 
  2M17511127-3406383 & 267.796969 & -34.110645  \\ 
  2M17523300-3027521 & 268.137518 & -30.464495 \\ 
  2M17534571-2949362 & 268.44047  & -29.826744 \\ 
  2M17552461-0122088 & 268.852559 & -1.369136 \\ 
  2M17554454-2123058 & 268.93562 & -21.384953 \\ 
  2M17555660-3238250 & 268.985848   & -32.640282 \\ 
  2M17560439-3246181 & 269.01833  & -32.771721  \\ 
  2M17571419-3328194 & 269.309144 & -33.472073 \\ 
  2M17573951-2908334 & 269.414629 & -29.142628 \\ 
  2M17595598-3117393 & 269.983287 & -31.294264 \\ 
  2M18013879-2924112 & 270.411633 & -29.403118 \\ 
  2M18014007-2649505 & 270.416966 & -26.830719 \\ 
  2M18014786-2749080 & 270.449436  & -27.818907 \\ 
  2M18015592-2749451 & 270.483011 & -27.829222 \\ 
  2M18033529-2911240 & 270.897062 & -29.19002 \\ 
  2M18035944-2908195 & 270.997669 & -29.138758 \\ 
  2M18044803-2752467 & 271.200154 & -27.879654  \\ 
  2M18050144-3005149 & 271.256017 & -30.087484 \\ 
  2M18054875-3122407 & 271.453164 & -31.377975 \\ 
  2M18061308-2522503 & 271.554505 & -25.380655 \\ 
  2M18062975-2855357 & 271.623993 & -28.926601 \\ 
  2M18072810-2459356 & 271.867096  & -24.993229 \\ 
  2M18100924-3733319 & 272.538504 & -37.55888 \\ 
  2M18101932-0930066 & 272.580527 & -9.50184 \\ 
  2M18120031-1350169 & 273.001326 & -13.838031  \\ 
  2M18121957-2926310 & 273.081553 & -29.441954 \\ 
  2M18315425-2328124 & 277.976045 & -23.470121 \\ 
  2M18334592-2903253 & 278.441366 & -29.057034  \\ 
  2M18360807-2314389 & 279.033649 & -23.244165 \\ 
  2M18364041-3402389 & 279.168375 & -34.044147 \\ 
  2M18425902-3007370 & 280.74595 & -30.126949 \\
  2M18442352-3029411 & 281.098036 & -30.494764 \\
  2M18475308-2602331 & 281.971167 & -26.042528 \\
  2M18562844-2814085 & 284.118507 & -28.23572 \\
  2M18594405-3651518 & 284.933562 & -36.864391 \\
  2M19175998-2919360 & 289.499952 & -29.326691 \\
\hline
\end{tabular}
\label{ids}
\end{table*}

\end{document}